\documentclass[nofootinbib, reprint, preprintnumbers, amsmath, amssymb, amsfonts, aps, pra, superscriptaddress, floatfix]{revtex4-2}
\usepackage[utf8]{inputenc}
\usepackage{mathtools}
\usepackage{graphicx}
\usepackage{dcolumn}
\usepackage{bm}
\usepackage{physics}
\usepackage{float}
\usepackage{placeins}
\usepackage{color}
\usepackage{braket}
\usepackage[normalem]{ulem}
\usepackage{enumerate}
\usepackage{enumitem}
\usepackage[colorlinks=true,linkcolor=blue,citecolor=blue,urlcolor =blue]{hyperref}
\usepackage{mathrsfs}
\usepackage{bbm}
\usepackage{siunitx}
\usepackage{graphicx,color}
\usepackage{amssymb}
\usepackage{amsmath}
\usepackage{placeins} 
\usepackage{subcaption}
\usepackage{bm}
\usepackage{ragged2e}
\usepackage{comment}
\usepackage{lipsum}

\begin{document}
\title{
Analysis of Superradiance-Based Quantum Metrology under Independent Markovian Pure Dephasing
}

\author{Yasunari Sakuma}
\affiliation{\mbox{School of Fundamental Science and Technology, Keio University, Yokohama, Kanagawa 223-8522, Japan}}%




  

\author{Yuichiro Matsuzaki}
\email{ymatsuzaki872@g.chuo-u.ac.jp}
\affiliation{Department of Electrical, Electronic, and Communication Engineering, Faculty of Advanced Science and Engineering, Chuo University, Kasuga, Bunkyo-ku, Tokyo 112-8551, Japan}

\author{Hoi-Kwan Lau}
\affiliation{Department of Physics, Simon Fraser University, Burnaby, British Columbia V5A 1S6, Canada}

\author{Junko Ishi-Hayase}
\email{hayase@appi.keio.ac.jp}
\affiliation{\mbox{School of Fundamental Science and Technology, Keio University, Yokohama, Kanagawa 223-8522, Japan}}
\affiliation{\mbox{Center for Spintronics Research Network, Keio University, Yokohama, Kanagawa 223-8522, Japan}}

\date{\today}


\begin{abstract}

Recently, a DC magnetometry protocol utilizing $N$-spin-ensemble superradiance was proposed. This method physically amplifies the acquired signal, suppressing estimation errors from measurement noise and achieving $\mathcal{O}(1/N)$ precision scaling when measurement noise dominates quantum fluctuations. However, quantum metrology is generally vulnerable to independent Markovian pure dephasing. For instance, the scaling of Greenberger-Horne-Zeilinger (GHZ) state-based magnetometry deteriorates from $\mathcal{O}(1/N)$ to $\mathcal{O}(1/\sqrt{N})$. Although pure dephasing likely degrades superradiant sensing, its quantitative impact remains unclear. Here, we investigate the effect of independent Markovian pure dephasing on this protocol using numerical simulations and mean-field analysis. We demonstrate that, in the large-$N$ limit, the estimation error increase is limited to a constant factor. This sharply contrasts with GHZ-state-based sensing, where the error increases by a factor of $\sqrt{N}$. Our analytical solutions elucidate the physical origin of this robustness qualitatively. These findings establish the high robustness of superradiance-based DC magnetometry against independent Markovian pure dephasing.
\end{abstract}

\maketitle
\section{Introduction}

Precise DC magnetometry plays a central role in a wide range of fields, from basic science to applied technologies. To achieve even higher precision, quantum metrology exploits powerful sensing protocols utilizing quantum resources. Quantum metrology protocols are typically evaluated by the estimation error, which quantifies the deviation between estimated and true values \cite{Budker2007}. When utilizing $N$ independent spins, the central limit theorem restricts the scaling of the estimation error to $\mathcal{O}(1/\sqrt{N})$ . In contrast, utilizing non-classical correlations, such as quantum entanglement, is known to improve this scaling to $\mathcal{O}(1/N)$ . This has stimulated active research from both theoretical and experimental perspectives \cite{GiovannettiRamseyinterfe,Spinsqueezing,Vittorio_Giovannetti2004,Jonathan_A2009}.

However, many entanglement-based protocols are vulnerable to environmental pure dephasing \cite{generalframework}. Particularly in DC magnetic field sensing, their advantages are significantly diminished in the presence of independent Markovian pure dephasing on each qubit\cite{generalframework,Demkowicz2012}. For instance, a DC magnetometry protocol that uses Greenberger–Horne–Zeilinger (GHZ) states as initial states is typical. Although this method exhibits some resistance to decoherence perpendicular to the quantization axis\cite{Brask2015,Chaves2013,Kessler2014,Dur2014,Arrad2014,Isogawa2023}, its estimation error scaling degrades from $\mathcal{O}(1/N)$  to $\mathcal{O}(1/\sqrt{N})$  when independent Markovian pure dephasing acts on each qubit \cite{Improvementoffrequency}. Under such pure dephasing conditions, entangled states can maintain their superiority over independent states only under specific circumstances, such as when the quantum noise is spatially correlated \cite{Jeske2014} or non-Markovian \cite{Matsuzaki2011,Chin2012}.

Recently, a new DC magnetometry method has been theoretically proposed that utilizes superradiance in a spin-1/2 ensemble for signal amplification \cite{WOS:000850727100001}. Superradiance is the phenomenon of collective decay in a qubit ensemble coupled to a lossy bosonic mode \cite{DICKERH}. The interaction mediated by a common bosonic mode causes the phases of the dipole moments of individual qubits to align, creating correlations in the emission process. Consequently, a coherent pulse is emitted with a peak intensity proportional to $N^2$ and a duration inversely proportional to $N$. This phenomenon has been experimentally realized in various setups, including atomic and molecular ensembles \cite{OBSERVATIONOFDICKESUPERRADIANCE,superradultracoldRydbergg,Gross1976}, superconducting qubits \cite{superradiancetwoartificialatoms}, quantum dots\cite{Scheibner2007} and quantum systems in diamond \cite{Superradiantemissionfromcolourentresindiamond}. Owing to the physical signal amplification by the superradiant process, this method can achieve a scaling of $\mathcal{O}(1/N)$  in regimes where measurement noise is dominant, although the scaling remains $\mathcal{O}(1/\sqrt{N})$  when quantum projection noise is the dominant factor.

However, existing proposals do not account for the effects of pure dephasing processes occurring during the signal acquisition phase. In superradiance-based quantum metrology, pure dephasing experienced during the interaction with a DC magnetic field is expected to negatively affect both the encoded phase information and the subsequent signal amplification process. Nevertheless, the extent to which such pure dephasing degrades the signal amplification factor, as well as its quantitative impact on the final estimation error, has remained unclear. As demonstrated by the example of GHZ states, independent Markovian pure dephasing can be a primary factor in the loss of quantum metrological advantages. Furthermore, pure dephasing is a ubiquitous and unavoidable noise source in many realistic quantum systems. Therefore, clarifying the robustness of superradiance-based methods against   pure dephasing represents an important challenge for the realization of practical quantum sensors. 

To address this challenge, in this study, we quantitatively evaluate the effect of independent Markovian pure dephasing on the signal amplification and the estimation error in superradiance-based quantum metrology, utilizing both an analytical approach based on the mean-field approximation and numerical simulations. As a result, we reveal that in the large-$N$ limit, the degradation of the estimation error remains merely a constant factor compared to the ideal case. This result is  in contrast to measurements utilizing GHZ states, which lose their scaling advantage under independent Markovian pure dephasing. 
The physical mechanism underlying this difference in robustness can be understood as follows: Magnetometry using GHZ states employs a highly entangled state for the entire system during the interaction with the magnetic field. Consequently, the total pure dephasing rate of the system increases to $N$ times that of a single spin, leading to an $N$-dependent increase in the estimation error. Conversely, superradiance-based magnetometry does not require quantum entanglement between spins during the sensing phase. Therefore, the pure dephasing probability per spin remains constant and independent of $N$. Furthermore, the analytical expressions derived in this study quantitatively demonstrate that almost all spins that escape pure dephasing during the interaction continues to contribute collectively to signal amplification. This confirms that the increase in estimation error is limited to a constant factor independent of $N$, demonstrating that superradiant quantum magnetometry is an effective method for enhancing sensitivity even in the presence of independent Markovian pure dephasing.

The remainder of this paper is organized as follows: Section II provides an overview of conventional DC magnetometry using Ramsey interferometry. Section III describes the theoretical model of the superradiance-based quantum metrology protocol proposed in previous studies~\cite{WOS:000850727100001}. In Section IV, we formulate the   pure dephasing process and present the results of our scaling analysis based on the mean-field approximation, supported by numerical simulations. Within this section, we also discuss the validity of the approximate model and the underlying physical mechanisms. Finally, Section V presents our conclusions.

\section{DC magnetometry using Ramsey interferometry}

\begin{figure}[htbp]
    \centering
    \includegraphics[width=0.95\columnwidth]{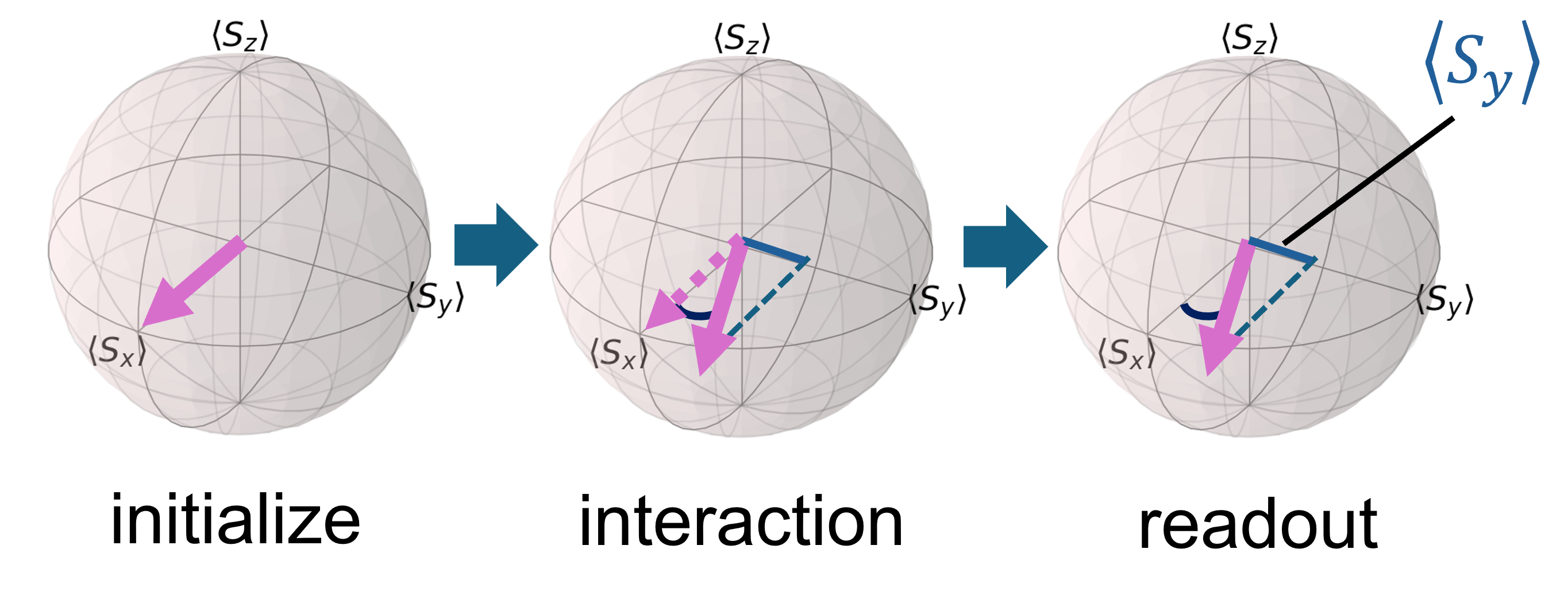}
    \caption{\justifying Schematic diagram of the protocol for DC magnetometry using Ramsey interferometry.$\ev{S_x}$, $\ev{S_y}$ and $\ev{S_z}$ represent the expectation values of the collective spin components along the $x$, $y$, and $z$ axes, respectively.}
    \label{fig:ramsey_protocol}
\end{figure}

In this section, we briefly explain a standard Ramsey interferometry protocol using an ensemble of independent two-level spins\cite{Improvementoffrequency}, which we will compare with the superradiance-based method discussed in subsequent sections.
The measurement protocol for Ramsey interferometry is illustrated in Fig.~\ref{fig:ramsey_protocol}.

The procedure is as follows. Initially, $N$ spins are prepared in a coherent spin state (CSS) pointing in the $x$-direction ($\otimes _{j=1}^N \ket{+}_j$, where $\hat{\sigma}_x\ket{+}_j=\ket{+}_j$) (initialize).
Next, the spins interact with a target DC magnetic field for an interaction time $\tau$, accumulating a relative phase of $\omega_0 \tau$ due to Larmor precession (interaction).
Here, the interaction Hamiltonian is given by $\hat{H}=\omega_0 \hat{S}_z= (\omega_0/2) \sum_{j=1}^N \hat{\sigma}_z^{(j)}$.
Finally, the magnetic field strength is estimated by reading out the expectation value of the magnetization in the $y$-direction, $\ev{S_y}= (1/2) \sum_{j=1}^N \ev{\sigma_y^{(j)}}$ (readout).

\begin{figure*}[tb]
    \includegraphics[width=0.75\textwidth]{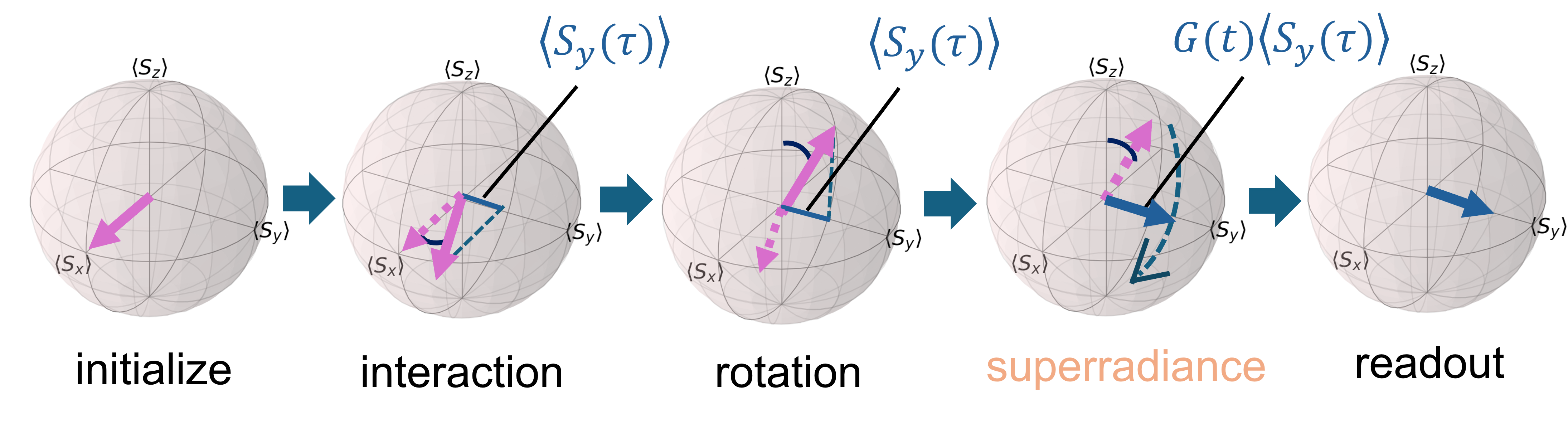}
\caption{\justifying Schematic diagram of the protocol for DC magnetometry using superradiance. $\ev{S_x}$, $\ev{S_y}$ and $\ev{S_z}$ represent the expectation values of the collective spin components along the $x$, $y$ and $z$ axes, respectively. In our model, we assume that Markovian pure dephasing primarily affects the 'interaction' and 'rotation' steps.}
\label{fig:SR_protocol}
\end{figure*}

The estimation error $\delta \omega$, which represents the uncertainty in the magnetic field estimation, arises from the statistical fluctuations of the measurement outcomes and is defined by the following equation \cite{Taylor2008,Budker2007}:
\begin{equation}
    \label{eq:ee}
    \delta \omega = \frac{\sqrt{C_{yy}}}{\sqrt{T/\tau} \abs{\pdv{\ev{{S}_y}}{\omega_0}}}
\end{equation}
Here, $\hat{S}_y$ is an observable to be measured, $C_{yy}= \ev{S^2_y} - \ev{S_y}^2$ is its variance, and $T$ is the total measurement time. We assume that the time required for initialization and readout is sufficiently short and can be neglected.

First, we evaluate the estimation error $\delta \omega_{\mathrm{d}}$ for the case without pure dephasing but with measurement noise. We assume a depolarizing noise model that acts immediately before an ideal measurement, degrading the density matrix $\hat{\rho}$ into a completely mixed state with probability $p$. This is represented by the map $\rho \to (1-p)\hat{\rho} + p \frac{\hat{I}}{2^N}$.
The estimation error $\delta \omega_{\mathrm{d}}$ is then given by 
\begin{align}
    \delta \omega_{\mathrm{d}} &= \frac{\sqrt{\ev{S_y^2} - \ev{S_y}^2}}{\sqrt{T/\tau} \abs{\pdv{\ev{S_y}}{\omega_0}}} \notag \\
    &\approx \sqrt{1 + \frac{p}{1-p}} \frac{1}{\sqrt{T \tau (1-p) N}} .\label{eq:nomal_ramsey}
\end{align}

Next, we consider the case where independent Markovian pure dephasing acts on each spin, in addition to the measurement noise. In this case, the time evolution of the density matrix $\hat{\rho}(t)$ of the spin ensemble during the interaction is described by the following Lindblad master equation:
\begin{equation}
    \label{eq:interaction_dephasing}
    \dv{\hat{\rho}(t)}{t}  = i\left[ \hat{\rho}(t),\omega_0 \hat{S}_z\right]
    + \sum_{i=1}^N \frac{\gamma}{2}\left(\hat{\sigma}_z^{(i)} \hat{\rho}(t) \hat{\sigma}_z^{(i)} - \hat{\rho}(t)\right).
\end{equation}
The first term on the right-hand side represents the unitary time evolution driven by the DC magnetic field, and the second term represents the effect of Markovian pure dephasing acting on each spin independently at a rate $\gamma$.
Under this time evolution, the expectation values of the spins at time $\tau$ decay due to the pure dephasing effects.
After the interacton with the magnetic filed, $\ev{{S}_y(\tau)}$ and $\ev{S^2_{y}(\tau)}$ are given by $\ev{S_y(\tau)} = \frac{N}{2} \sin(\omega_0 \tau) \exp(-\gamma \tau)$,\,$\ev{S^2_y(\tau)} = \frac{N}{4}+ \frac{N(N-1)}{4}\exp(-2\gamma \tau )\sin^2(\omega_0 \tau)$ respectively.
Substituting these into Eq.~\eqref{eq:ee}, we obtain the estimation error $\delta \omega_{\mathrm{sep}}$ under pure dephasing. Here, we assume that the accumulated phase is sufficiently small ($N(\omega_0\tau)^2 \ll 1$), yielding:
\begin{equation}
    \label{eq:relax_nomal_ramsey} 
    \delta \omega_{\mathrm{sep}} \approx \sqrt{1 + \frac{p}{1-p}} \frac{1}{\sqrt{T \tau (1-p) N} \exp(-\gamma \tau)}.
\end{equation}
Equations~\eqref{eq:nomal_ramsey} and \eqref{eq:relax_nomal_ramsey} show that the estimation error scales as $\mathcal{O}(1/\sqrt{N})$ with respect to the number of spins $N$.

Finally, we determine optimal interaction time $\tau$ that minimize the estimation error$\delta \omega_{\mathrm{sep}}$.
By differentiating Eq.~\eqref{eq:relax_nomal_ramsey} with respect to $\tau$, the condition $\pdv{\delta \omega_{\mathrm{sep}}}{\tau} = 0$ reveals that estimation error reaches its minimum at 
\begin{equation}
    \gamma \tau = \frac{1}{2},
\end{equation}
independent of measurement noise parameter $p$.

\section{DC magnetometry using superradiance}

In this section, we outline the theoretical model for DC magnetic field sensing using superradiance, building upon a previous study \cite{WOS:000850727100001} that forms the foundation of our work. This method physically amplifies a weak phase signal ($\ev{{S}_y}$) via the phenomenon of superradiance. By pre-amplifying the signal itself prior to measurement, the signal-to-noise ratio is enhanced, thereby reducing the final estimation error.

To describe the superradiant dynamics in this study, we employ the following Lindblad master equation \cite{GROSSMandHAROCHES}:
\begin{equation}
    \label{eq:master}
    \dv{t} \hat{\rho}_s(t) = \Gamma \left( \hat{S}_-\hat{\rho}_s(t)\hat{S}_+ - \frac{1}{2} \{\hat{S}_+\hat{S}_-, \hat{\rho}_s(t)\} \right).
\end{equation}

Here, $\hat{\rho}_s(t)$ is the density matrix of the spin ensemble, and $\Gamma$ represents the spontaneous relaxation rate of a single spin. The operators $\hat{S}_{\pm} = \sum_{i=1}^N \hat{\sigma}_{\pm}^{(i)}$ are the collective spin raising and lowering operators, and $\{\cdot, \cdot\}$ denotes the anticommutator.

The measurement protocol for superradiance-based DC magnetometry is illustrated in Fig.~\ref{fig:SR_protocol}.
First, similar to the Ramsey interferometry described in the previous section, an initial spin-coherent state aligned along the $x$-axis is prepared and interacts with the target DC magnetic field for an interaction time $\tau$. Next, a $\pi/2$ pulse is applied to the system along the $y$-axis to drive the ensemble into a nearly fully excited state (we set this time to $t = 0$,). The signal component ($\ev{{S}_y(\tau)}=\frac{N}{2}\sin(\omega_0 \tau)$) is then physically amplified by the ensuing superradiance. This superradiant emission can be controlled by adjusting the detuning and the coupling between the resonator and the spin ensemble. Finally, the amplified signal is read out.
The essence of this method lies in amplifying only the amplitude of the signal while preserving the accumulated initial phase information $\ev{{S}_y(\tau)}$. By applying a mean-field approximation \cite{WOS:000850727100001,GENERALIZEDCUMULANT} that sets specific covariances to zero for the time evolution of $\ev{{S}_y(t)}$ and $\ev{{S}_z(t)}$ derived from the master equation \eqref{eq:master}, we obtain the following coupled equations:
\begin{align}
    \label{eq:MFz}\dv{t}\ev{{S}_z(t)} &= \Gamma \ev{{S}_z(t)}(\ev{{S}_z(t)}-1) -\Gamma C^{\prime}, \\
    \label{eq:MFy}\dv{t}\ev{{S}_y(t)} &= \Gamma \ev{{S}_y(t)}\left(\ev{{S}_z(t)} - \frac{1}{2}\right),\\
    \label{eq:MFyy}\dv{C_{yy}}{t} = \Gamma& [ ( 2\ev{S_z}C_{yy} - 1) + \ev{S_z(t)}^2 -\frac{1}{2} \ev{S_z(t)}],
\end{align}
where $t \ge \tau$, $C_{yy}\equiv \ev{{S}^2_y(t)}-\ev{{S}_y(t)}^2$, and $C^{\prime} \equiv \frac{N}{2}\left(\frac{N}{2}+1\right)$ represents the expectation value of the total angular momentum. 
Assuming a small phase rotation ($N(\omega_0 \tau)^2 \ll 1$), $\ev{{S}_y(\tau)}$ takes a small but non-zero value. Under this condition, taking the state immediately after the $\pi/2$ pulse as the initial value, i.e., $\ev{{S}_y(0)}=\ev{{S}_y(\tau)}=\frac{N}{2}\sin(\omega_0 \tau)$, the time evolution of $\ev{S_y(t)}$ can be solved as follows\cite{WOS:000850727100001}:
\begin{equation}
    \label{eq:sy_amp}
    \ev{{S}_y(t)} \approx \frac{\cosh{\left(\frac{1}{2}\log{N}\right)}}{\cosh{\left(\frac{N}{2}\Gamma t - \frac{1}{2}\log{N}\right)}}\ev{{S}_y(0)}.
\end{equation}
Equation \eqref{eq:sy_amp} clearly demonstrates that the signal $\ev{{S}_y(t)}$ is amplified in the form of $G(t)\ev{{S}_y(0)}$ as time evolves, where $G(t)\equiv \ev{{S}_y(t)}/\ev{{S}_y(0)}$. This gain $G(t)$ scales up to $\sqrt{N}$ at optimal time $t_0 \approx \ln(N)/N\Gamma$. This $N$-dependent signal amplification ultimately reduces the estimation error.
When $N$ is sufficiently large, the maximum value of $G(t)$, denoted as $G_{\text{max}}$, is given by
\begin{align}
    \label{eq:g_max}
    G_{\text{Max}} &= \frac{1}{2} \qty( \sqrt{N} + \frac{1}{\sqrt{N}} ) \approx \frac{\sqrt{N}}{2}.
\end{align}
Thus, this amplification factor is proportional to $\sqrt{N}$ at its maximum. This $N$-dependent signal amplification reduces the estimation error.

Next, we evaluate the impact of this signal amplification on the estimation error, $\delta \omega$, in the presence of measurement noise. 
Incorporating this amplification effect, the estimation error is calculated based on the maximum gain $G_{\text{Max}}$ derived from Eq.~\eqref{eq:sy_amp}, yielding the following expression:
\begin{align}
  \delta \omega_{\mathrm{SR}}
    \label{eq:nodepasing_err} &= \sqrt{(1+\sigma^2_{\mathrm{add}}) + \frac{p}{(1-p)c^2_0N}} \frac{1}{\sqrt{aN}}.
\end{align}
Here, $\sigma^2_{\mathrm{add}}$ is a phenomenological coefficient representing the spin fluctuations along the $y$-axis that are amplified during superradiance. The parameters $c_0$ and $a$ are dimensionless factors defined by $G_{\text{Max}} = c_0\sqrt{N}$ and $a=T \tau(1-p)$, respectively.

Inside the square root of Eq. \eqref{eq:nodepasing_err}, the first term represents the noise originating from quantum fluctuations \cite{WOS:000850727100001}, while the second term accounts for the contribution of the measurement noise. In the asymptotic limit of $N \to \infty$, the quantum fluctuation term becomes dominant, and the error asymptotically approaches $\mathcal{O}(1/\sqrt{N})$. Crucially, however, the second term associated with the measurement noise is suppressed by a factor of $\mathcal{O}(1/N)$. Consequently, in the regime where the measurement noise is dominant, the estimation error can temporarily drop below the $\mathcal{O}(1/\sqrt{N})$ scaling.
It should be noted, however, that this behavior does not imply the attainment of the Heisenberg limit because initial probe is unentangled.

\section{Results under independent Markovian pure dephasing}
In this section, we present our findings regarding the impact of independent Markovian pure dephasing on the estimation error in superradiance-based DC magnetometry. 
For this analysis, we employ two approaches: an analytical method based on the mean-field approximation and numerical simulations using the Lindblad master equation.

\begin{figure*}[t]  
    \centering
    \captionsetup{justification=centering}
    \begin{minipage}[b]{0.45\textwidth}
        \centering
        \includegraphics[width=\linewidth]{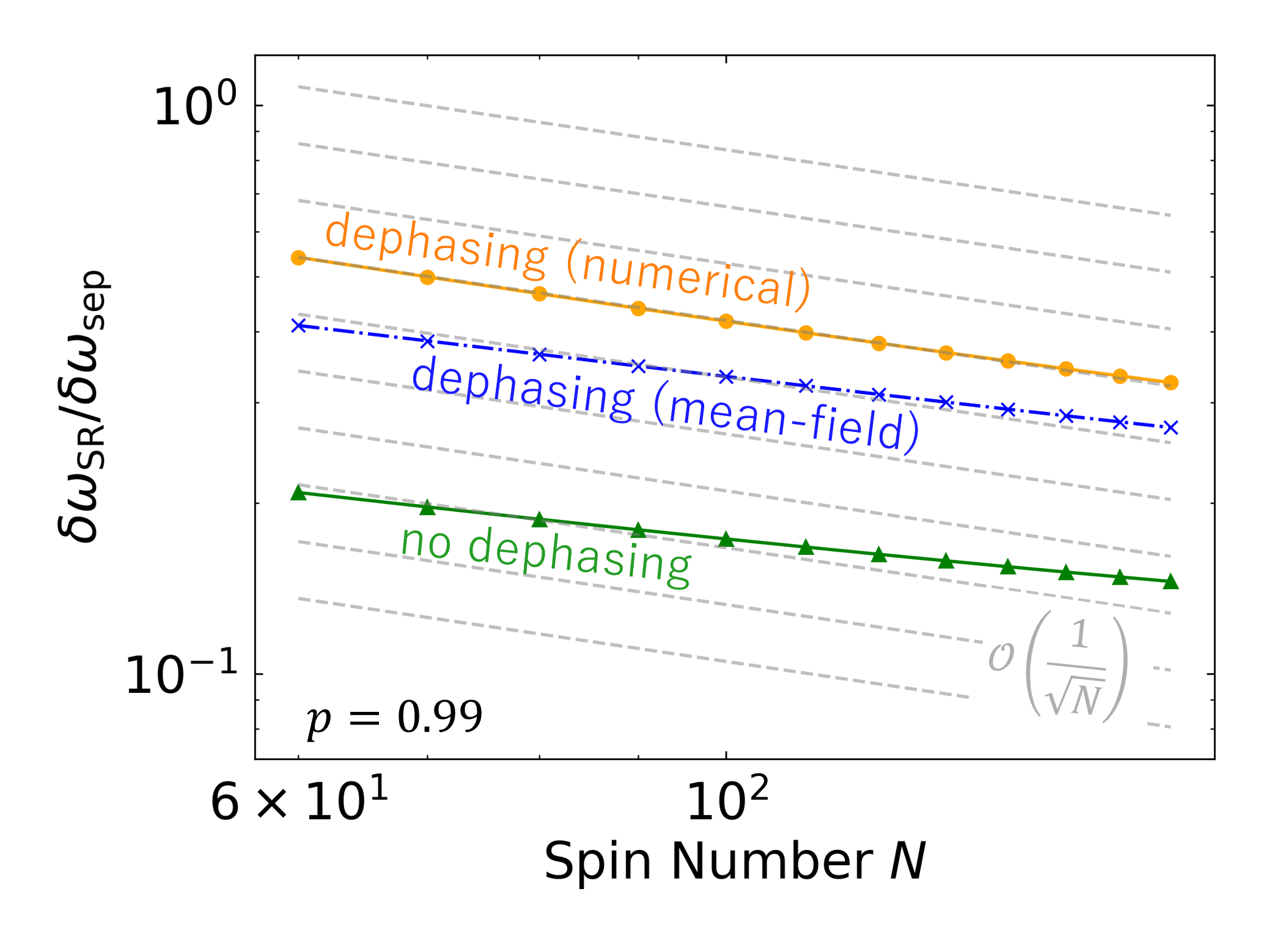} \\
        (a) $p = 0.99$
    \end{minipage}
    \hfill
    \begin{minipage}[b]{0.45\textwidth}
        \centering
        \includegraphics[width=\linewidth]{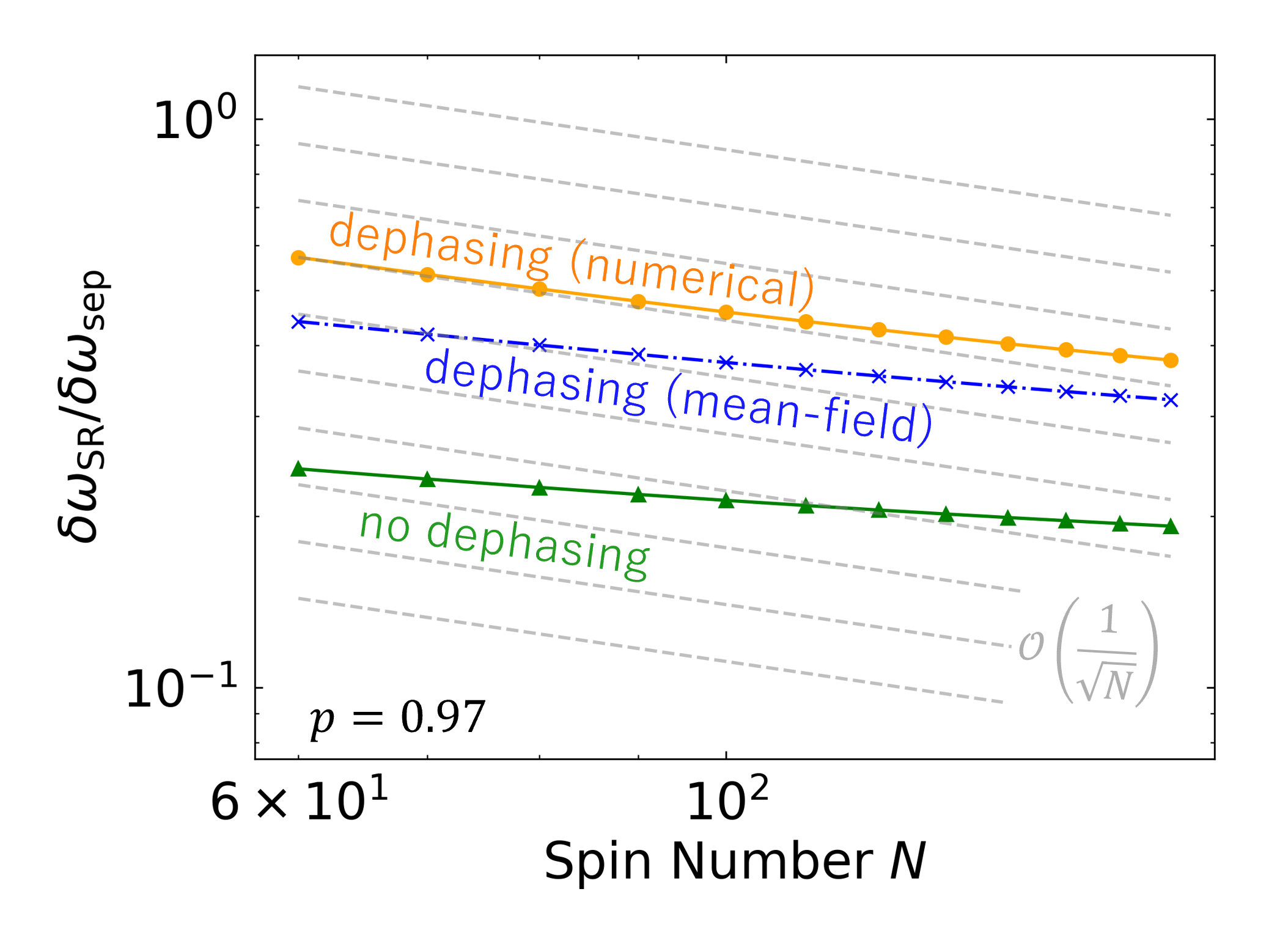} \\
        (b) $p = 0.97$
    \end{minipage}
   \caption{\justifying Dependence of the estimation error ratio $\delta\omega_{\text{SR}}/\delta\omega_{\text{sep}}$ on the number of spins $N$. For the simulation results (orange line and green line), the point that numerically minimizes the estimation error is selected. The parameters are taken from Table 1: coupling strength $g = 10^4~\mathrm{Hz}$, cavity decay rate $\kappa = 10^6~\mathrm{Hz}$, and pure dephasing rate $\gamma = 1~\mathrm{kHz}$.}
    \label{fig:error_scaling_combined_wide}
\end{figure*}

\subsection{Scaling Analysis of Signal Amplification Factor and Estimation Error via Mean-Field Approximation}

We consider a system in which independent, Markovian pure dephasing acts only during the interaction time ($\tau$) with the magnetic field, and depolarizing noise with probability $p$ is introduced during the final readout, as illustrated in Fig.~\ref{fig:SR_protocol}.
In other words, we neglect the pure dephasing during the $\pi/2$ rotation pulse and the superradiance process, since their timescales($\sim 1/N\Gamma$) are considered much shorter than the pure dephasing timescale($\sim1/\gamma$). Under these approximations, the time evolution of the density matrix $\hat{\rho}(t)$ of the spin ensemble during the interaction is described by Eq.~\eqref{eq:interaction_dephasing}.
\begin{align}
    C &= \ev{S_x(0)^2} + \ev{S_y(0)^2} + \ev{S_z(0)^2} \notag \\
      &= \frac{N^2}{4}\exp(-2\gamma \tau) + \frac{N}{4}\left(3-\exp(-2\gamma \tau)\right), \\
      \ev{S_z(0)}&=\frac{N}{2}\exp(-\gamma \tau)\cos(\omega_0 \tau),\\
    \ev{S_y(0)} &= \frac{N}{2}\exp(-\gamma \tau)\sin(\omega_0 \tau), \\
    \ev{S^2_y(0)} &= \frac{N}{4}+ \frac{N(N-1)}{4}\exp(-2\gamma \tau )\sin^2(\omega_0 \tau).
\end{align}

From Eqs.~\eqref{eq:MFz}--\eqref{eq:MFyy}, we obtain the signal amplification factor $G(t)=\ev{S_y(t)}/\ev{S_y(0)}$ at time $t \ge 0$ following the onset of superradiance. Here, the constant $C^{\prime}$ in Eq.~\eqref{eq:MFz} is replaced by $C$. $G(t)$ is expressed as follows:
\begin{align}
    \label{eq:Gt_exact}
    G(t) &= \frac{\cosh{\left( \Gamma\sqrt{C+\frac{1}{4}}\,t_0 \right)}}
    {\cosh{\left( \Gamma\sqrt{C+\frac{1}{4}}\,(t-t_0) \right)}}, \\
    \label{eq:t0_def}
    t_0 &= \frac{1}{\Gamma \sqrt{1+4C}}
   \ln\left( \frac
             {\sqrt{\frac{1}{4}+C}+\ev{S_z(0)}-\frac{1}{2}}{\sqrt{\frac{1}{4}+C}-\ev{S_z(0)}+\frac{1}{2}} \right).
\end{align}
The maximum amplification factor is given by $G_{\text{Max}} \equiv G(t_0) = \cosh\left(\Gamma\sqrt{C+\frac{1}{4}}\,t_0\right)$ at time $t=t_0$.

Now, we consider the case where the system satisfies the condition $1 \ll N \exp(-\gamma\tau)$. The quantity $N \exp(-\gamma\tau)$ can be interpreted as the effective number of spins that remain undephased, relative to the initial number of qubits $N$. When this value is sufficiently larger than 1, $G_{\text{Max}}$ can be approximated as follows:
\begin{align}
    G_{\text{Max}} 
    &=
     \sqrt{\frac{Ne^{-\gamma \tau} + \frac{1}{Ne^{-\gamma \tau}} +3e^{\gamma \tau} -e^{-\gamma \tau} }{(3-e^{-\gamma \tau})(e^{\gamma \tau}+1)}} \\
     &\approx  \sqrt{\frac{Ne^{-\gamma \tau} }{(3-e^{-\gamma \tau})(e^{\gamma \tau}+1)}}
     .
    \label{eq:Gmax_approx}
\end{align}
Comparing this expression with Eq.~\eqref{eq:g_max}, which represents the case without dephasing, reveals that the $N$-dependent term is replaced by $Ne^{-\gamma \tau}$. This $Ne^{-\gamma \tau}$ term can be interpreted as the number of spins that escape pure dephasing during the interaction, and the expression demonstrates that this effective number contributes almost directly to the signal amplification.

Evaluating the estimation error $\delta\omega_{\text{SR}}$ using this maximum amplification factor yields the following:
\begin{widetext}
\begin{align}
\delta\omega_{\text{SR}}
&=\label{eq:error_SR_exact}\sqrt{\frac{4}{N G_{\text{Max}}^2}(C_1 - \frac{1}{4}) + \frac{p}{(1-p)G_{\text{Max}}^2}} \,\frac{1}{ \sqrt{T\tau}\sqrt{N(1-p)}e^{-\gamma \tau}}, \\
&\label{eq:error_SR_approx}\approx \sqrt{(1+e^{-\gamma \tau}) + \frac{(3-e^{-\gamma \tau})(e^{\gamma \tau}+1)p}{(1-p)Ne^{-\gamma \tau}}} \,\frac{1}{ \sqrt{T\tau}\sqrt{N(1-p)}e^{-\gamma \tau}} ,
\end{align}
\end{widetext}
where $C_1$ is given by
\begin{align}
C_1 &\approx \frac{N^2e^{-2\gamma \tau}}{4(3-e^{-\gamma \tau})}.
\end{align}

Comparing this expression with the term proportional to $1/N$ in Eq.~\eqref{eq:nodepasing_err}, which represents the effect of suppressing detector noise, we see that the factor $N$ has been replaced by $Ne^{-\gamma\tau}$. This implies that, much like in the expression for signal amplification, the spins that do not undergo pure dephasing contribute almost directly to the reduction of the estimation error.

Equation~\eqref{eq:error_SR_approx} is an approximate solution obtained by retaining only the leading-order terms with respect to $N$ in $C_1$ and $G_{\text{Max}}$.
Comparing the second term inside the square root, which corresponds to the suppression of detector noise, with the ideal-case term in Eq.~\eqref{eq:nodepasing_err} yields
\begin{align}
    \frac{\dfrac{p(3-e^{-\gamma \tau})(1+e^{\gamma \tau})}{(1-p)e^{-\gamma \tau}N}}{\dfrac{p}{(1-p)c_0^2N}}
    = c_0^2(3e^{\gamma \tau} -1)(e^{\gamma \tau}+1),
\end{align}
which is a constant independent of $N$ and $p$. This implies that even under the influence of independent Markovian pure dephasing, the degradation of the estimation error is limited to a constant factor relative to the ideal case.

Furthermore, evaluating $G_{\text{Max}}$ under the condition of a large pure dephasing rate $\gamma$ yields
\begin{align}
    G_{\text{Max}} 
    &\approx \sqrt{\frac{2 +3e^{\gamma \tau} -e^{-\gamma \tau} }{(3-e^{-\gamma \tau})(e^{\gamma \tau}+1)}} = 1 \notag \\
    &\hspace{3cm} \text{for}\ Ne^{-\gamma \tau} \sim 1, \\
    G_{\text{Max}}
    &\approx \sqrt{\frac{ 1}{Ne^{-\gamma \tau}(3-e^{-\gamma \tau})(e^{\gamma \tau}+1)}} \notag \\
    &\hspace{3cm} \text{for}\ Ne^{-\gamma \tau} \ll 1.
\end{align}
These results indicate that the amplification factor no longer scales with $N$, meaning that scaling advantage is lost even when measurement noise dominates quantum fluctuations.
These considerations reveal that to preserve the favorable scaling, the superradiance-based noise suppression technique requires a regime satisfying at least $1 \ll N \exp(-\gamma\tau)$. This corresponds to a state where a sufficient number of spins retain their coherence. 
In the experiment, this condition can be achieved by employing a sufficiently large $N$ and optimizing the magnetic field interaction time $\tau$.

\begin{figure*}[t]
    \centering
    \captionsetup{justification=centering}
    \begin{minipage}[b]{0.32\textwidth}
        \centering
        \includegraphics[width=\linewidth]{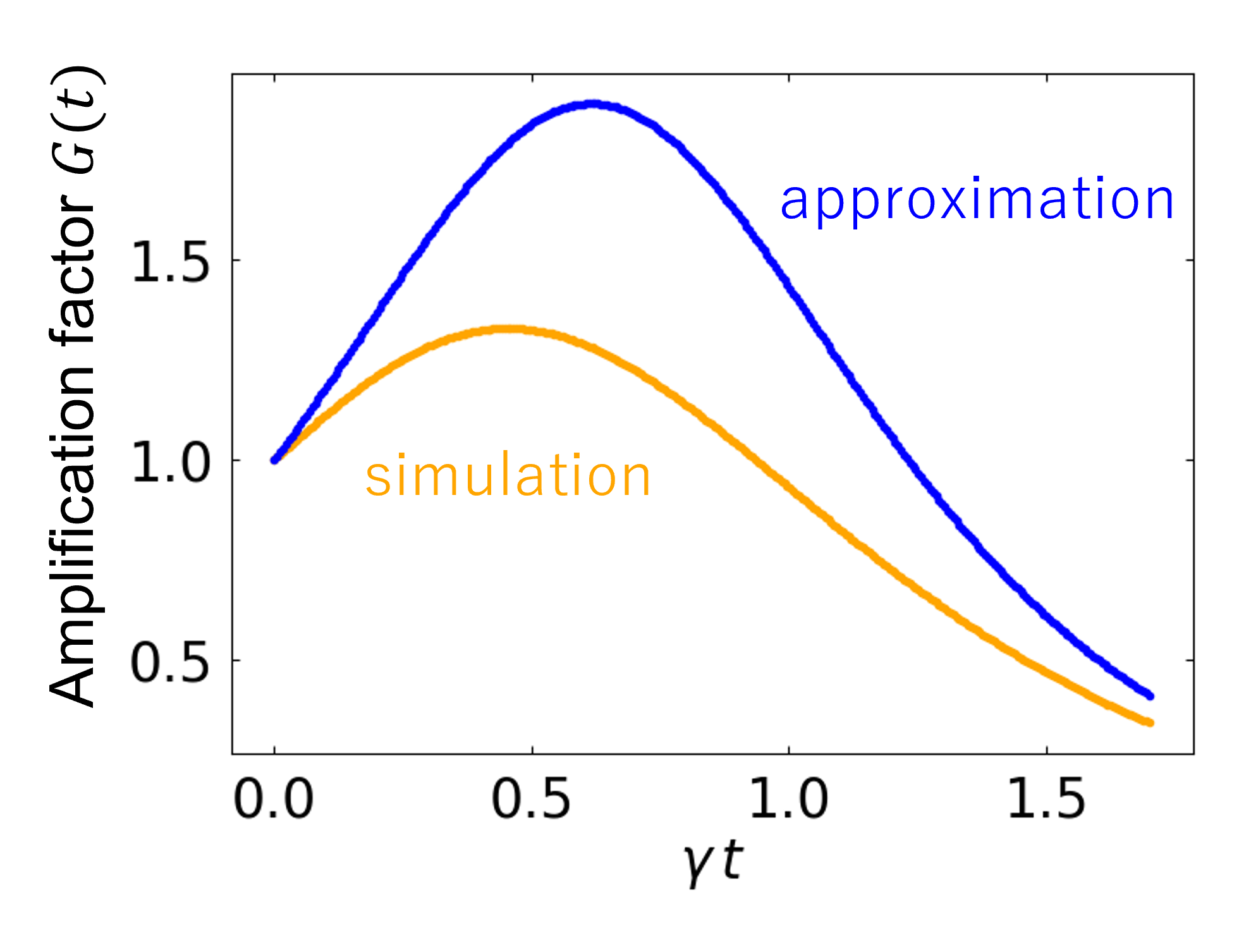} \\
        (a) $N=30$
        \label{fig:6.2.1}
    \end{minipage}
    \hfill
    \begin{minipage}[b]{0.32\textwidth}
        \centering
        \includegraphics[width=\linewidth]{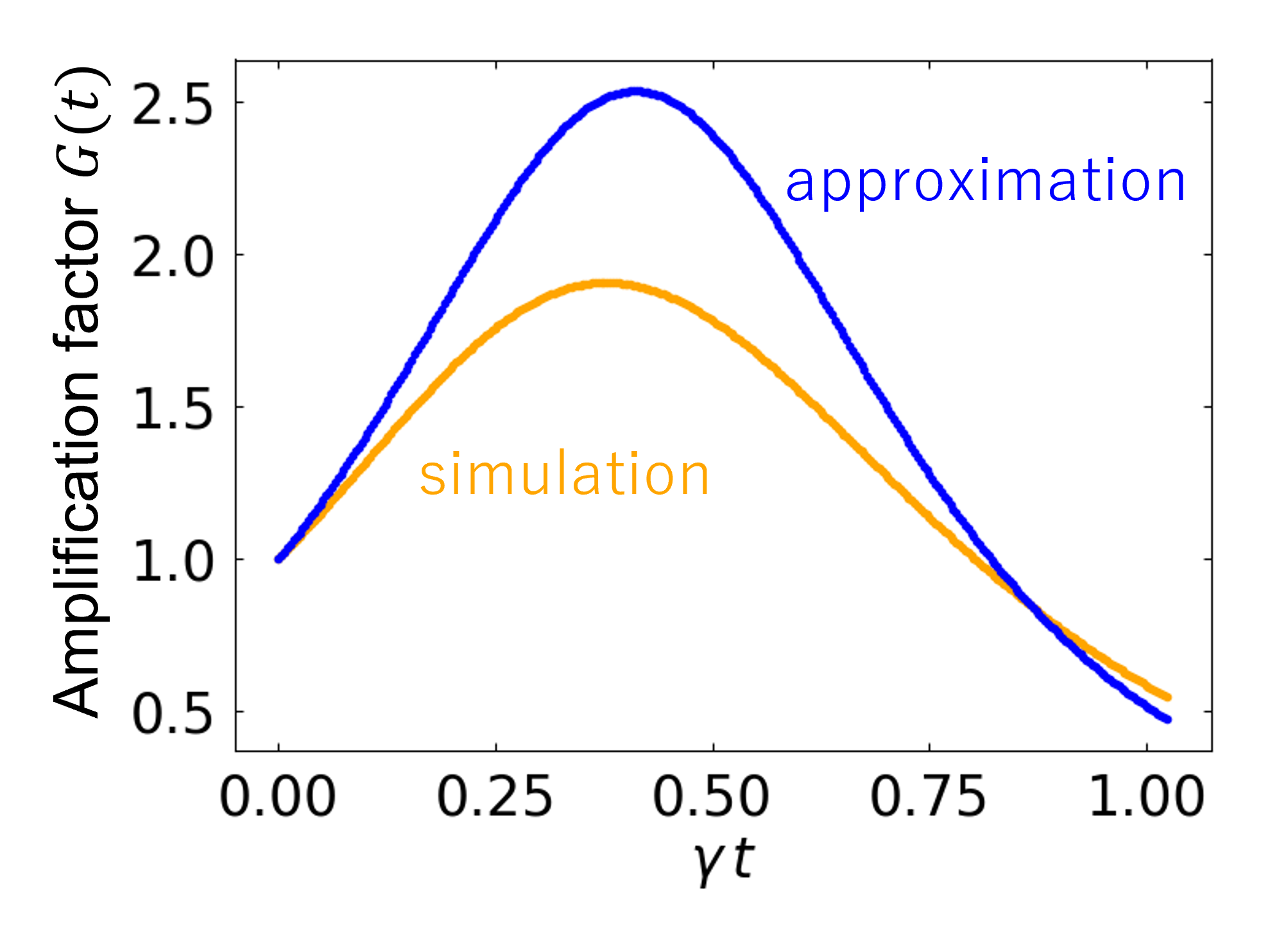} \\
        (b) $N=60$
        \label{fig:6.2.2}
    \end{minipage}
    \hfill
    \begin{minipage}[b]{0.32\textwidth}
        \centering
        \includegraphics[width=\linewidth]{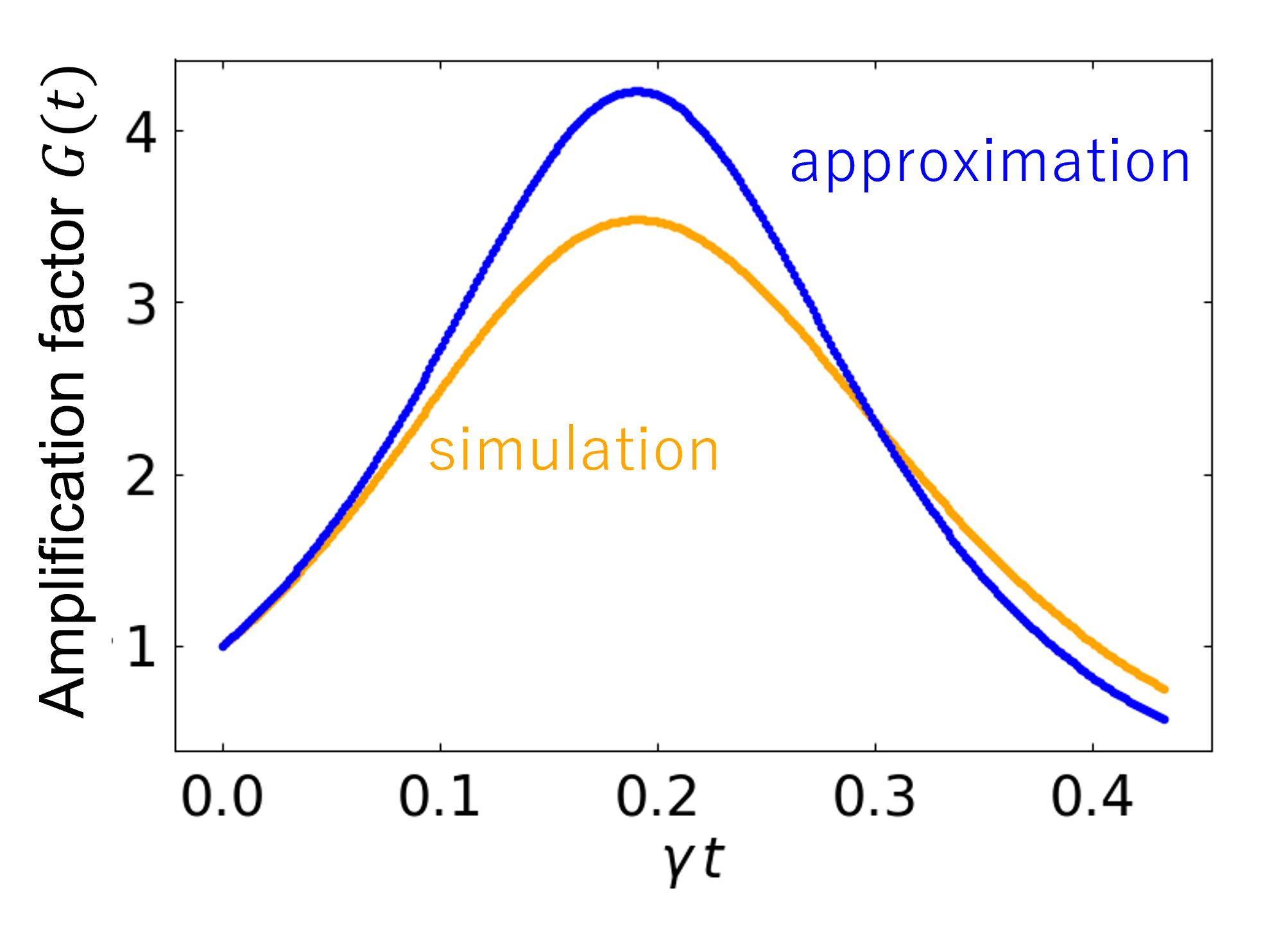} \\
        (c) $N=180$
        \label{fig:6.2.3}
    \end{minipage}
\caption{\justifying Comparison the amplification factor between the numerical simulation (orange) and the approximate solution (blue) for each number of spins $N$ at the interaction time $\gamma \tau = 1/2$. The parameters are taken from Table 1: coupling strength $g = 10^4~\mathrm{Hz}$, cavity decay rate $\kappa = 10^6~\mathrm{Hz}$, and pure dephasing rate $\gamma = 1~\mathrm{kHz}$.}
    \label{fig:combined_N}
\end{figure*}

\subsection{Scaling Analysis using Numerical Simulations}

In this section, we investigate the scaling of the estimation error $\delta\omega_{\text{SR}}$ through numerical simulations of the Lindblad master equation. In contrast to the mean-field analysis presented earlier, our simulations adopt a more realistic framework in which independent Markovian pure dephasing continuously affects the system throughout the entire magnetometry protocol. This includes not only the interaction time but also the $\pi/2$ rotation pulses and the superradiant emission dynamics. Additionally, our model considers a spin ensemble coupled to a cavity, where superradiance is triggered by the Purcell effect. Specifically, the superradiance rate is controlled by adjusting the detuning between the spin ensemble and the cavity resonance. The single-spin relaxation rate $\Gamma$ induced by the Purcell effect is given by $\Gamma=\frac{4g^2}{\kappa}$, where $g$ denotes the coupling strength and $\kappa$ is the cavity decay rate \cite{KoshinoKandShimizuA}.

Table~\ref{tab:simulation_params} summarizes the parameters employed in the simulations. As a representative realistic solid-state quantum system, we adopted experimental values from erbium-ion ($\mathrm{Er^{3+}}$) spins doped in a calcium tungstate ($\mathrm{CaWO_4}$) crystal \cite{Single-electronspin}. Because this system features a relatively strong spin-cavity coupling ($g \sim 10^4~\mathrm{Hz}$), it adequately captures the rapid dynamics inherent in superradiance even under the computational limitations associated with simulating a small number of spins. This makes it an ideal model for our study.

\begin{table}[H]
    \centering
    \caption{Parameters used for the numerical simulations \cite{Single-electronspin}}
    \label{tab:simulation_params}
    \begin{tabular}{lcc} \hline \hline
        Parameter & Symbol & Typical Value \\ \hline
        Coupling strength & $g$ & $10^4~\mathrm{Hz}$ \\ 
        Cavity decay rate & $\kappa$ & $10^6~\mathrm{Hz}$ \\ 
        Dephasing rate & $\gamma$ & $1/T_2 \sim 1~\mathrm{kHz}$ 
       \\ \hline \hline
    \end{tabular}
\end{table}

To reflect realistic experimental conditions, we specify the value of the depolarization parameter $p$ introduced in the previous section. For example, in  nitrogen-vacancy (NV) center in diamond, a solid-state quantum sensing platforms, the measurement uncertainty is known to increase by a factor of $10$ to $10^3$ compared to the ideal projection noise limit, primarily due to low photon collection efficiency and limited readout contrast \cite{Sensitivityoptimization}. As formalized in Eq.~\eqref{eq:nomal_ramsey}, the parameter $p$ characterizes these imperfections, leading to an increase in the estimation error by a factor of $1/(1-p)$. Consequently, setting $p = 0.9$--$0.999$ allows us to phenomenologically capture such realistic measurement degradations within our simulation framework.




The numerical simulations were implemented using the Permutational Invariant Quantum Solver (PIQS) module in QuTiP \cite{QuTiP:Anopen-sourcePython,QuTiP2:APythonframework,permutational_invariance,Kirton2017}. This solver takes advantage of the permutational invariance of the spin ensemble to significantly reduce the dimensionality of the Hilbert space, thereby enabling the efficient computation of large-scale many-body dynamics. Based on the discussion in the previous section, the interaction time $\tau$ with the DC magnetic field is fixed at $\gamma \tau = 1/2$, which is the optimal value for Ramsey interference, to facilitate a direct comparison.

We investigated the dependence of the estimation error $\delta\omega_{\text{SR}}$ on the number of spins $N$. Figure~\ref{fig:error_scaling_combined_wide} presents the ratio of the superradiant estimation error $\delta\omega_{\text{SR}}$ to the standard Ramsey estimation error $\delta\omega_{\text{sep}}$ for two distinct noise parameters: (a) $p=0.99$ and (b) $p=0.97$. Note that $\delta\omega_{\text{sep}}$ is calculated assuming exposure to pure dephasing during the interaction time in Fig.~\ref{fig:ramsey_protocol}.
Since $\delta\omega_{\text{sep}}$ is proportional to $1/\sqrt{N}$, if this ratio $\delta\omega_{\text{SR}}/\delta\omega_{\text{sep}}$ scales as $\mathcal{O}(1/\sqrt{N})$ (i.e., follows the dashed line in Fig.~\ref{fig:error_scaling_combined_wide}), it indicates that $\delta\omega_{\text{SR}}$ scales as $\mathcal{O}(1/N)$. On the other hand, if the ratio becomes constant with respect to $N$, it means that $\delta\omega_{\text{SR}}$ decreases on the order of $\mathcal{O}(1/\sqrt{N})$.
As illustrated in Fig.~\ref{fig:error_scaling_combined_wide}(a), the ratio for the case without pure dephasing (green line) monotonically decreases as $N$ increases, demonstrating a scaling advantage surpassing $\mathcal{O}(1/\sqrt{N})$. Crucially, at least within the parameter regime of our proposed scheme, this decreasing trend with respect to $N$ persists even in the presence of pure dephasing (orange line). This finding verifies that a temporary scaling advantage is sustained even in an independent Markovian pure dephasing environment.

Moreover, examining the simulation results for $p=0.97$ [Fig.~\ref{fig:error_scaling_combined_wide}(b), orange line], a scaling of $\mathcal{O}(1/N)$ is observed in the regime of relatively small $N$. Conversely, as $N$ grows, the scaling asymptotically reverts to $\mathcal{O}(1/\sqrt{N})$.This behavior is consistent with the trend reported in Ref.~\cite{WOS:000850727100001}, where the scaling becomes $\mathcal{O}(1/N)$ in the measurement-noise-dominated regime and returns to $\mathcal{O}(1/\sqrt{N})$ in the quantum-fluctuation-dominated regime. Our result shows that this trend occurs even under  pure dephasing.

\subsection{Comparison between Numerical Simulations and Approximate Solutions}

As shown in Fig.~\ref{fig:error_scaling_combined_wide}, the analytical solution obtained using the mean-field approximation (blue line) qualitatively agrees with the numerical simulation results; however, it tends to overestimate the amplification factor. This discrepancy is attributed to the neglect of the variance of the spin operators in the mean-field approximation.

From the definition of variance, the inequality $\ev{S^2_z(t)}-\ev{S_z(t)}^2 \geq 0$ always holds. However, since the mean-field approximation assumes $\ev{S^2_z(t)}-\ev{S_z(t)}^2 \approx 0$ in its derivation, $\ev{S_z(t)}$, which corresponds to the population of the excited state, is overestimated. Consequently, the effective number of spins contributing to the superradiant process is also overestimated, resulting in a higher amplification factor for the signal intensity $\ev{S_y(t)}$ than that obtained from the numerical simulation. This corresponds to the overestimation on the right-hand side of Eq.~\eqref{eq:MFy}.

Furthermore, Fig.~\ref{fig:error_scaling_combined_wide}(b) shows that the slope of the approximate solution flattens at an earlier stage compared to the numerical simulation. This is also due to the mean-field approximation overestimating the maximum amplification factor $G_{\text{Max}}$. Because $G_{\text{Max}}$ is overestimated, the error caused by measurement noise is more rapidly suppressed. This leads to an earlier transition to the regime where quantum fluctuations dominate, which exhibits a slope of $\order{1/\sqrt{N}}$. As a result, the approximate curve flattens earlier than the simulation curve, which maintains a continuous $\order{1/N}$ decrease over a wider range.

As seen in Fig.~\ref{fig:error_scaling_combined_wide}, the approximate solution approaches the numerical simulation results as the number of spins $N$ increases. This is considered to be because the validity of our approximations in the large $N$ limit. To verify this, Fig.~\ref{fig:combined_N} shows the time evolution of the amplification factor for different numbers of spins ($N=30, 60$ and $180$) with an interaction time of $\gamma \tau=1/2$.

Although the approximate solution still overestimates the amplification factor, it increasingly agrees with the overall dynamics of the simulation results as $N$ increases. Here, we define the relative error $\epsilon_{\mathrm{rel}}$ for the maximum value of the amplification factor as follows:

\begin{equation}
    \epsilon_{\mathrm{rel}} = \abs{ \frac{G_{\mathrm{app}} - G_{\mathrm{sim}}}{G_{\mathrm{sim}}} }
    \label{eq:relative_error}
\end{equation}

where $G_{\mathrm{app}}$ and $G_{\mathrm{sim}}$ are the maximum amplification factors in the approximate solution and the numerical simulation, respectively. We found that while the relative error is $\epsilon_{\mathrm{rel}} \approx 0.329$ for $N=60$, it decreases to $\epsilon_{\mathrm{rel}} \approx 0.215$ for $N=180$. This quantitatively confirms that the accuracy of the approximation improves with increasing $N$.

\subsection{Origin of Robustness against Independent Markovian pure dephasing}

The above analysis shows that in the quantum metrology method using superradiance, even in the presence of independent Markovian pure dephasing during the interaction, the degradation in the scalings of both the signal amplification factor and the estimation error is limited to a constant factor. This is in contrast to the metrology method utilizing GHZ states, where the estimation error increases by a factor of $\order{\sqrt{N}}$ under the same conditions. The physical origin of this difference in robustness can be summarized as follows.

In metrology using GHZ states, because the entire spin system is in a strongly entangled state from the beginning of the interaction stage, the pure dephasing rate of the whole system increases in proportion to the number of spins $N$. Consequently, the effective interaction time available for signal accumulation is limited to $1/N$, leading to the degradation of the scaling of the final estimation error to $\order{1/\sqrt{N}}$~\cite{Improvementoffrequency}.

On the other hand, in superradiant quantum metrology, an unentangled state where spins are independent of each other is used during the interaction. Therefore, the number of spins that maintain coherence after the interaction time $\tau$ is given by $Ne^{-\gamma\tau}$. According to Eq.~\eqref{eq:Gmax_approx}, which is the analytical solution obtained using the mean-field approximation in this study, the subsequent scaling of the signal amplification factor due to superradiance quantitatively depends on the number of spins that survived pure dephasing during the interaction, $Ne^{-\gamma\tau}$. 
The qualitative reason for this is that superradiance only requires the coupling between single spins and a common electromagnetic field, and does not require special states such as spin entanglement in the initial state.
This result indicates that the spins that escaped pure dephasing contribute almost directly to the collective signal amplification. Therefore, even if some spins lose coherence due to pure dephasing, the system can maintain a macroscopic signal amplification factor of $\mathcal{O}(\sqrt{N})$ as long as the number of spins that have not undergone pure dephasing remains on the order of $\mathcal{O}(N)$. Because this amplification factor of $\mathcal{O}(\sqrt{N})$ effectively suppresses detector noise and directly contributes to improving the scaling of the estimation error, we conclude that superradiant quantum metrology possesses high robustness against pure dephasing.

\section*{Conclusions}

In conclusion, we theoretically investigated the effects of independent Markovian pure dephasing on the scalings of the signal amplification factor and the estimation error in superradiance-based DC magnetometry. Specifically, we derived an analytical solution using the mean-field approximation and performed numerical simulations based on the Lindblad master equation, verifying the consistency between the two approaches.

Our analysis revealed that even in the presence of pure dephasing, the increase in the estimation error in superradiance-based DC magnetometry is limited to a constant factor compared to the dephasing-free case. 

Furthermore, we discussed the physical mechanism behind this robustness. Superradiance-based metrology does not require quantum entanglement between spins during the interaction. Therefore, the probability of an individual spin undergoing pure dephasing remains constant regardless of the total number of spins $N$. Moreover, our derived approximate solution indicates that the spins remaining unaffected by pure dephasing during the interaction, $Ne^{-\gamma \tau}$, continue to contribute almost directly to the signal amplification.
Intuitively, this is because superradiance requires only the coupling between multiple individual spins and a common electromagnetic field, without requiring special initial states such as spin-spin entanglement.
Consequently, the increase in the estimation error is suppressed to a constant factor independent of $N$. These results demonstrate that the signal amplification scheme via superradiance possesses high robustness against independent Markovian pure dephasing.

\section*{Acknowledgment}
This work is partly supported by MEXT Q-LEAP (Grant No. JPMXS0118067395
), JSPS KAKENHI (Grant No. JP25K01292
), JST CREST (Grant No. JPMJCR24A5, Grant No. JPMJCR23I5), JST PRESTO (Grant No. JPMJPR245B
), SIP program, Program for the Advancement of
Next Generation Research Projects, Keio University and Center for Spintronics Research Network.

\appendix
\section{
Derivation of the amplification gain and estimation error
}
In this appendix, we derive Eqs.~\eqref{eq:Gt_exact} and \eqref{eq:error_SR_approx}. To solve the time evolution of $\ev{S_y}$ and its fluctuation $\ev{S_y^2}$ in the superradiant dynamics under the mean-field approximation, we must first determine the initial state of the superradiance process. Specifically, we require the conserved quantity $C$ related to the total angular momentum, as well as the initial values $\ev{S_y(0)}$ and $\ev{S_z(0)}$ at the onset of superradiance (i.e., at the end of the interaction period). 
Using Eq.~\eqref{eq:interaction_dephasing}, these are given as follows:
\begin{align}
    C &= \ev{S_x(0)^2} + \ev{S_y(0)^2} + \ev{S_z(0)^2} \notag \\
      &= \frac{N^2}{4}\exp(-2\gamma \tau) + \frac{N}{4}\left(3-\exp(-2\gamma \tau)\right), \\
    \ev{S_z(0)} &= \frac{N}{2}\exp(-\gamma \tau)\cos(\omega_0 \tau), \\
    \ev{S_y(0)} &=\ev{S_y(\tau)}= \frac{N}{2}\exp(-\gamma \tau)\sin(\omega_0 \tau), \\
    \ev{S_y^2(0)} &= \frac{N}{4} + \frac{N(N-1)}{4}\exp(-2\gamma \tau)\sin^2(\omega_0 \tau).
\end{align}
Throughout this paper, we assume a small phase rotation ($N(\omega_0 \tau)^2 \ll 1$).

Using these initial values, we evaluate the signal gain $G(t) = \ev{S_y(t)}/\ev{S_y(\tau)}$ after a time $t$ has elapsed from the start of superradiance. To this end, we solve Eqs.~\eqref{eq:MFz}, \eqref{eq:MFy}, and \eqref{eq:MFyy}. Hereafter, time-dependent expectation values such as $\ev{S_y(t)}$ and $\ev{S_z(t)}$ are abbreviated as $\ev{S_y}$ and $\ev{S_z}$, respectively.
\begin{align}
    \tag{\ref{eq:MFz}}\dv{t}\ev{S_z} &= \Gamma \ev{S_z}(\ev{S_z}-1) - \Gamma C, \\
    \tag{\ref{eq:MFy}}\dv{t}\ev{S_y} &= \Gamma \ev{S_y}\left(\ev{S_z} - \frac{1}{2}\right), \\
    \tag{\ref{eq:MFyy}}\dv{C_{yy}}{t} &= \Gamma \left[ -(1 - 2\ev{S_z})C_{yy} + \ev{S_z}^2 - \frac{1}{2}\ev{S_z} \right].
\end{align}

First, we solve the differential equation given by Eq.~\eqref{eq:MFz}.
By considering the roots of the quadratic equation
\begin{equation}
    x^2-x-C=0,
\end{equation}
which are $x_{\pm}=\frac{1}{2}(1\pm \sqrt{1+4C})$, we can rewrite Eq.~\eqref{eq:MFz} as follows:
\begin{equation}
    \frac{ d{\ev{S_z}}/d{t}}{\left(\ev{S_z}-x_+\right)\left(\ev{S_z}-x_-\right)}
   =\Gamma.
\end{equation}
Applying partial fraction decomposition and integrating both sides with respect to time yields
\begin{equation}
    \ln{\left( \frac{\sqrt{\frac{1}{4}+C}-\ev{S_z}+\frac{1}{2}}{\sqrt{\frac{1}{4}+C}+\ev{S_z}-\frac{1}{2}} \right)}
    =\Gamma \sqrt{1+4C}\,t  + A,
\end{equation}
where $A$ is the constant of integration.
By rearranging these expressions, we obtain the analytical solution for $\ev{S_z}$:
\begin{equation}
    \label{eq:newMFz}
    \ev{S_z} = -\sqrt{C+\frac{1}{4}} \tanh{\left[ \frac{\Gamma}{2}\sqrt{1+4C}\,(t-t_0) \right]}+\frac{1}{2},
\end{equation}
with $t_0$ defined as
\begin{equation}
      \label{eq:re_t0_def}
   t_0=\frac{1}{\Gamma\sqrt{1+4C}}
   \ln{\left( \frac{\sqrt{\frac{1}{4}+C}+\ev{S_z(0)}-\frac{1}{2}}{\sqrt{\frac{1}{4}+C}-\ev{S_z(0)}+\frac{1}{2}} \right)}.
\end{equation}

Next, to solve Eq.~\eqref{eq:MFy}, we rewrite it as follows:
\begin{align}
    \frac{1}{\ev{S_y}}
    \dv{\ev{S_y}}{t} =
   \Gamma(\ev{S_z}-\frac{1}{2}), \notag\\
 \label{A_eq11}\Leftrightarrow\quad
  \dv{\ln{\ev{S_y}}}{t} =
   \Gamma(\ev{S_z}-\frac{1}{2}).
\end{align}
Substituting Eq.~\eqref{eq:newMFz} into Eq.~\eqref{A_eq11} and integrating both sides, we obtain
\begin{align*}
\ev{S_y}&=\ev{S_y(\tau)}\exp[\int_{0}^tdt^\prime
    ( \Gamma(\ev{S_z}-\frac{1}{2}))],
    \\&=
    \ev{S_y(\tau)}\exp[\int_{0}^tdt^\prime
    \Gamma
     (-\sqrt{C+\frac{1}{4}}
    \tanh{\frac{\Gamma}{2}\sqrt{1+4C}\,(t^\prime-t_0)})],
    \\&
    =\ev{S_y(\tau)}\exp[-\int_{-\Gamma\sqrt{C+\frac{1}{4}}\,t_0}
    ^{\Gamma\sqrt{C+\frac{1}{4}}\,(t-t_0)}
    dq\tanh{q}],
    \\&=
    \ev{S_y(\tau)}
    \frac{\cosh{\Gamma\sqrt{C+\frac{1}{4}}\,t_0}}
    {\cosh{\Gamma\sqrt{C+\frac{1}{4}}\,(t-t_0)}}.
\end{align*}
From these equations, we derive Eq.~\eqref{eq:Gt_exact}. The expression for the amplification factor $G(t)$ is given by
\begin{align}
    \label{eq:re_Gt_exact_ap}
    G(t) &= \frac{\cosh{\left( \Gamma\sqrt{C+\frac{1}{4}}\,t_0 \right)}}
    {\cosh{\left( \Gamma\sqrt{C+\frac{1}{4}}\,(t-t_0) \right)}}.
\end{align}

The amplification factor reaches its maximum value, $G_{\text{Max}}$, at time $t=t_0$. Assuming $1 \ll N \exp(-\gamma\tau)$, we approximate $G_{\text{Max}}$ to derive Eq.~\eqref{eq:Gmax_approx} as follows:
\begin{widetext}
\begin{align}
    G_{\text{Max}} 
    =
     \sqrt{\frac{Ne^{-\gamma \tau} + \frac{1}{Ne^{-\gamma \tau}} +3e^{\gamma \tau} -e^{-\gamma \tau} }{(3-e^{-\gamma \tau})(e^{\gamma \tau}+1)}} 
     \approx  \sqrt{\frac{Ne^{-\gamma \tau} }{(3-e^{-\gamma \tau})(e^{\gamma \tau}+1)}}
     .
\end{align}
\end{widetext}
During superradiance, the quantity $C_{yy}=\ev{S^2_y}-\ev{S_y}^2$ is determined from Eq.~\eqref{eq:MFyy} as
\begin{align}
    \dv{C_{yy}}{t} = \Gamma [ -(1 - 2\ev{S_z})C_{yy}
+ \ev{S_z}^2 -\frac{1}{2} \ev{S_z}].\notag
\end{align}
Using Eq. \eqref{eq:newMFz} and introducing the variables $B=\sqrt{C+\frac{1}{4}}$ and $u=\Gamma B(t-t_0)$, this can be rewritten as
\begin{align}
    \dv{C_{yy}}{u} +2 \tanh(u)C_{yy} = \tanh(u) (B\tanh(u)-\frac{1}{2}).\notag
\end{align}
From this, we find
\begin{equation}
    \dv{(C_{yy}\cosh^2(u))}{u} =
    B\sinh^2(u)-\frac{1}{4}\sinh(2u),
\end{equation}
which yields the solution:
\begin{widetext}
\begin{align}
    C_{yy}&=\frac{C_1}{\cosh^2(u)} + \frac{B}{2}\tanh(u) -\frac{Bu}{2\cosh^2(u)} - \frac{1}{4},\\
    C_1 &= \frac{N+1}{8}  +\frac{N+1}{16} \left( \frac{\sqrt{\frac{1}{4}+C}-\ev{S_z(\tau)}+\frac{1}{2}}
{\sqrt{\frac{1}{4}+C}+\ev{S_z(\tau)}-\frac{1}{2}} + \frac
{\sqrt{\frac{1}{4}+C}+\ev{S_z(\tau)}-\frac{1}{2}}{\sqrt{\frac{1}{4}+C}-\ev{S_z(\tau)}+\frac{1}{2}} \right)  \notag \\
&+\frac{\sqrt{\frac{1}{4}+C}}{8}
\left( \frac
{\sqrt{\frac{1}{4}+C}+\ev{S_z(\tau)}-\frac{1}{2}}{\sqrt{\frac{1}{4}+C}-\ev{S_z(\tau)}+\frac{1}{2}}
- \frac{\sqrt{\frac{1}{4}+C}-\ev{S_z(\tau)}+\frac{1}{2}}
{\sqrt{\frac{1}{4}+C}+\ev{S_z(\tau)}-\frac{1}{2}}  \right)\notag + \frac{\sqrt{\frac{1}{4}+C}}{4}\ln\left( \frac{\sqrt{\frac{1}{4}+C}-\ev{S_z(\tau)}+\frac{1}{2}}
{\sqrt{\frac{1}{4}+C}+\ev{S_z(\tau)}-\frac{1}{2}} \right).
\end{align}
\end{widetext}

Next, we evaluate the leading-order term of $C_{yy}$ at $t=t_0$, where the amplification factor reaches its maximum. Noting that $C_{yy}(t=t_0) = C_1 - 1/4$, we extract the leading-order term of $C_1$ under the condition $1 \ll N \exp(-\gamma\tau)$:
\begin{align}
    C_1 &\approx \frac{N^2e^{-2\gamma \tau}}{4(3-e^{-\gamma \tau})}.
\end{align}
Finally, we derive $\delta \omega_{\text{SR}}$ as follows:
\begin{widetext}
\begin{align}
\delta\omega_{\text{SR}}
&=\frac{\sqrt{\ev{S_y^2} - \ev{S_y}^2}} {\sqrt{T/\tau}\pdv*{\ev{S_y}}{\omega_0}}\notag 
\approx \frac{\sqrt{ Np/4 + (1-p)[ C_{yy}+ pG_{\text{Max}}^2\ev{S_y(\tau)}^2 ]}}
    {\sqrt{T/\tau}(1-p)G_{\text{Max}}\pdv*{\ev{S_y(\tau)}}{\omega_0}} \notag \\
&=  \sqrt{\frac{4}{N G_{\text{Max}}^2}\left(C_1 - \frac{1}{4}\right) + \frac{p}{(1-p)G_{\text{Max}}^2}} \,\frac{1}{ \sqrt{T\tau}\sqrt{N(1-p)}e^{-\gamma \tau}}\\
\label{eq:re_error_SR_approx}
&\approx\sqrt{(1+e^{-\gamma \tau}) + \frac{(3-e^{-\gamma \tau})(1+e^{\gamma \tau})p}{(1-p)Ne^{-\gamma \tau} }} \,\frac{1}{ \sqrt{T\tau}\sqrt{N(1-p)}e^{-\gamma \tau}}.
\end{align}
\end{widetext}
Here, Eq.~\eqref{eq:re_error_SR_approx} reproduces Eq.~\eqref{eq:error_SR_approx} from the main text, which is the leading-order approximation with respect to $G_{\text{Max}}$ and $C_{yy}$.

\bibliography{ref}

\end{document}